\documentclass{aa}

\usepackage{graphicx}
\usepackage{natbib}
\usepackage{txfonts}

\begin{document}

\title {Effects of Galactic Fountains and delayed mixing in the chemical evolution of the Milky Way }

\author {E. Spitoni\inst{1} \thanks {email to: spitoni@oats.inaf.it}
  \and F. Matteucci\inst{1,2} \and S. Recchi\inst{3,2} \and
  G. Cescutti\inst{1} \and A. Pipino \inst{4}} \institute{
  Dipartimento di Astronomia, Universit\`a di Trieste, via
  G.B. Tiepolo 11, I-34131, Trieste, Italy \and I.N.A.F. Osservatorio Astronomico di
  Trieste, via G.B. Tiepolo 11, I-34131, Trieste, Italy \and Institute of Astronomy,
  Vienna University, T\"urkenschanzstrasse 17, A-1180, Vienna,
  Austria \and Department of Physics and Astronomy, University of
  Southern California, Los Angeles, CA 90089-0484, U.S.A}

\date{Received xxxx / Accepted xxxx}

 \abstract{The majority of galactic chemical evolution models assumes
   the instantaneous mixing approximation (IMA). This assumption is
   probably not realistic as indicated by the existence of chemical
   inhomogeneities, although current chemical evolution models of the
   Milky Way can reproduce the majority of the observational
   constraints under the IMA.} {The aim of this paper is to test
   whether relaxing this approximation in a detailed chemical
   evolution model can improve or worsen the agreement with
   observations.  To do that, we investigated two possible causes for
   relaxing of the instantaneous mixing: i) the ``galactic
   fountain time delay effect'' and ii) the ``metal cooling time delay
   effect''.}  {We considered various galactic fountain time delays
   for the chemical enrichment from massive stars. 
    We then tested a
   time delay in the enrichment from stars of all masses due to gas
   cooling  in the range 0.5-1 Gyr.}  {We found
   that the effect of galactic fountains is negligible if an average
   time delay of 0.1 Gyr, as suggested in a previous paper, is assumed. 
   Longer time delays produce
   differences in the results but they are not realistic. We also
   found that the O abundance gradient in the disk is not affected by
   galactic fountains. The metal cooling time delays produce strong
   effects on the evolution of the chemical abundances only if we
   adopt stellar yields depending on metallicity. If instead,
   the yields computed for to the solar chemical composition are adopted,
   negligible effects are produced, as in the case of the galactic
   fountain delay.} 
    {The relaxation of the IMA by means of the
   galactic fountain model, where the delay is considered only for
   massive stars and only in the disk, does not affect the chemical
   evolution results. The combination of metal dependent yields and
   time delay in the chemical enrichment from all stars starting from
   the halo phase, instead, produces results at variance with
   observations.}


\titlerunning{Effects of Galactic Fountains and delayed mixing }
\authorrunning{Spitoni et al.}
\maketitle
 
\section{Introduction}

   The first attempts to follow the chemical evolution of galaxies
  (Schmidt 1963, Searle \& Sargent 1963, Tinsley 1974) were based on
  the {\it Simple Model} assumptions: 1) the system is one-zone and
  closed, namely there are no inflows or outflows, 2) the initial gas
  is primordial (no metals), 3) the initial mass function (IMF) is
  constant in time, 4) the gas is well mixed at any time ({\it
    instantaneous mixing approximation or IMA}), 5) stars more massive than 1 M$_\odot$ die
  instantaneously, while stars smaller than 1 M$_\odot$ live forever ({\it
    instantaneous recycling approximation} or IRA) . While the IRA
  has been relaxed in most of the recent chemical evolution models,
  there have been only a few attempts to relax the IMA mainly because of difficulties in
  estimating the dispersal and mixing time of the chemical elements
  (Roy \& Kunth 1995) and because models which retain IMA provide
  nevertheless an excellent fit to the data in the Milky Way
  (e.g. Fran\c{c}ois et al. 2004, hereafter F04).  Although there is no
doubt that IMA is not realistic (Schmidt 1963; Tinsley 1975), the
validity of this assumption depends on the time-scale of the mixing
processes. Malinie et al.(1993) claimed that due to chemical
inhomogeneities in the disk, mixing and star formation might be
delayed by $10^{8-9}$ yr.

 One of the very few studies in which IMA has been relaxed is the one
 of Thomas et al. (1998). They proposed a model in which 
   relaxing IMA was included by means of the splitting of the gas
 components into two different phases: the inactive and active
 gas phases. The first one consists in a hot and not homogeneously
 distributed component; stars cannot form out of this phase. The
 active phase is instead assumed to be cool and well mixed. They
 discussed various time-scales for the mixing of the stellar ejecta,
 and they found that a delay of the order of $10^{8}$ years leads to a
 better fit of the observational data in the [Mg/Fe]-[Fe/H] diagram, than in the IMA hypothesis.

 In this paper
  we test the relaxation of IMA in the detailed chemical
  evolution model of F04, considering two physically motivated time
  delays: i) the ``galactic fountain time delay'' effect and ii) the
  delay produced by the cooling of the metals (``metal cooling delay model'').

 Type II SNe usually
occur in OB associations containing several dozen massive stars.
Sequential supernova explosions create supershells which can break out of 
a stratified medium, producing bipolar outflows. The gas of the
supershells can fragment into clouds which eventually fall toward the
disk producing the  so-called galactic fountains. Spitoni et al. (2008) (hereafter Paper I) 
followed the clouds ejected by SN explosions in the Galaxy both in the
purely ballistic model (the cloud is subject only to the gravitational
potential of the Galaxy) and in the hybrid one where viscous
interactions between the extra-planar halo gas and the cloud are taken
into account.  They found that the range of the cloud orbits is quite
small. The clouds are generally directed outwards but the average
landing coordinates differ from the throwing coordinates by at most
$\sim$ 1 kpc. Only for a throwing coordinate of 12 kpc and an OB
association made of 500 SNe the ballistic model predicts a landing
coordinate $\sim$ 2 kpc larger than the throwing one. This result is
consistent with the works of Bregman (1980), Fraternali \& Binney
(2008) and Melioli et al. (2008a). Melioli et al. (2008b) presented
results for a multiple fountain model produced by randomly clustered
explosions of SNe originating in stellar associations spread over a
disk area of 8 kpc $^{2}$. As in the case of a single fountain, the
spreading of the SN ejecta back into the disk is not very large. Most of
the gas lifted up by fountains falls back within a distance $\Delta
$R=$\pm$0.5 kpc. Therefore it is unlikely that galactic fountains can
affect abundance gradients on large scales.

  However, as shown in Paper I, the fountains take a finite and
  non-negligible time to orbit around and fall back into the
  Galaxy. This implies a delay in the mixing of metals in the
  interstellar medium (ISM), which conflicts with the IMA
  assumption. Referring to the results of Paper I, in this paper we
  test the effect on the chemical evolution of different values of the
  delay originated by sequential SN explosions at different
  galactocentric distances.

In this work we also investigate another possible cause of delayed
mixing of the ISM such as the delay due to the fact that stars form
only in cold gas and that hot gas takes a finite time to cool.  
  However, there is still some debate in the literature on the
  cooling time-scale. Recchi et al. (2001) studied
  the effect of a single, instantaneous star-burst on the dynamical
  and chemical evolution of a gas-rich dwarf galaxy. They found that
  most of the metals reside in the cold gas phase already after a few
  tens of Myr. This result is mainly due to the assumed low SNII
  heating efficiency, and justifies the generally adopted homogeneous
  and instantaneous mixing of gas in chemical evolution models. On the
  other hand, Rieschick \& Hensler (2000) presented a chemodynamical
  model of the evolution of a dwarf galaxy in which most of the metals
  undergo a cycle lasting almost 1 Gyr before being incorporated into 
  the ISM. A similar result has been obtained by Tenorio-Tagle
  (1996). More recently, in the work of Harfst et al. (2006) a multi-phase
  description of the ISM was implemented: the ISM was treated as two
  dynamically independent gas phases: a cloudy and a diffuse one. The
  two gas phases are linked by the processes of
 condensation/evaporation and a drag force. In this model, it is assumed
 that stars form in
clouds and that clouds are destroyed by stellar feedback, thereby
self-regulating the star formation. Each cloud is assumed to form
stars only {\it after a time of inactivity}, because not all the gas in
clouds is dense or unstable enough for immediate star formation. In this work 
the time of inactivity considered is a few hundred Myr.

 Here, we simply 
implemented our ``metal cooling delay model'' in the detailed chemical
evolution model of F04 by means of a delay in the chemical enrichment.

 The paper is organized as follows: in  Sect. 2 we describe the
 reference model of F04, in  Sects. 3 and 4 we report the
 nucleosynthesis  prescriptions and the observational data,
 respectively. In Sect. 5 we present our galactic fountains delay
 model, and in  Sect. 6 the ``metal cooling delay model''. In Sect. 7 we
 report and discuss our results. Finally, we draw the main conclusions
 in  Sect. 8.

\section{The chemical evolution model for the Milky Way}

In the model of F04, the Galaxy is assumed to have formed by means of
two main infall episodes: the first forms the halo and the thick disk,
the second the thin disk.  The time-scale for the formation of the
halo-thick disk is 0.8 Gyr and the entire formation period does not
last more than 2 Gyr. The time-scale for the thin disk is much longer,
7 Gyr in the solar vicinity, implying that the infalling gas forming
the thin disk comes mainly from the intergalactic medium and not only
from the halo (Chiappini et al. 1997).  Moreover, the formation
timescale of the thin disk is assumed to be a function of the
galactocentric distance, leading to an inside out scenario for the
Galaxy disk build-up (Matteucci \& Fran\c cois 1989).  The galactic
thin disk is approximated by several independent rings, 2 kpc wide,
without exchange of matter between them.

 The main characteristic of the two-infall model is an almost
 independent evolution between the halo and the thin disk (see also
 Pagel \& Tautvaisienne 1995).  A threshold gas density of
 $7M_{\odot}pc^{-2}$ in the star formation process (Kennicutt 1989,
 1998, Martin \& Kennicutt 2001) is also adopted for the disk.

 The model well reproduces already an extended set of observational
 constraints. Some of the
 most important observational constraints are represented by the
 various relations between the abundances of metals
 (C,N,O,$\alpha$-elements, iron peak elements) as functions of the
 [Fe/H] abundance (see Chiappini et al. 2003a, b and F04) and by the
 G-dwarf metallicity distribution.  It is worth mentioning here that,
 although this model is probably not unique, in this respect it reproduces the
 majority of the observed features of the Milky Way.  Many of the
 assumptions of the model are shared by other authors (e.g. Prantzos
 \& Boissier 2000, Alib\'es et al. 2001, Chang et al. 1999).

The equation below describes the time evolution of $G_{i}$, namely the
mass fraction of the element $i$ in the gas:

\begin{displaymath}
\dot{G_{i}}(r,t)=-\psi(r,t)X_{i}(r,t)
\end{displaymath}
\smallskip
\begin{displaymath}
+\int\limits^{M_{Bm}}_{M_{L}}\psi(r,t-\tau_{m})Q_{mi}(t-\tau_{m})\phi(m)dm
\end{displaymath}
\begin{displaymath}
+A_{Ia}\int\limits^{M_{BM}}_{M_{Bm}}\phi(M_{B})\cdot\left[\int\limits_{\mu_{m}}^{0.5}f(\mu)\psi(r,t-\tau_{m2})Q^{SNIa}_{mi}(t-\tau_{m2})d\mu\right]dM_{B}
\end{displaymath}
\begin{displaymath}
+(1-A_{Ia})\int\limits^{M_{BM}}_{M_{Bm}}\psi(r,t-\tau_{m})Q_{mi}(t-\tau_{m})\phi(m)dm
\end{displaymath}
\begin{displaymath}
+\int\limits^{M_{U}}_{M_{BM}}\psi(r,t-\tau_{m})Q_{mi}(t-\tau_{m})\phi(m)dm
\end{displaymath}
\smallskip
\begin{equation}
+X_{A_{i}}A(r,t)
\end{equation}
where $X_{i}(r,t)$ is the abundance by mass of the element $i$ and
$Q_{mi}$ indicates the fraction of mass restored by a star of mass $m$
in form of the element $i$, the so-called ``production matrix'' as
originally defined by Talbot \& Arnett (1973). We indicate with
$M_{L}$ the lightest mass which contributes to the chemical enrichment
and it is set at $0.8M_{\odot}$; the upper mass limit, $M_{U}$, is set
at $100M_{\odot}$.

The star formation rate (SFR) $\psi(r,t)$ is defined as:
\begin{equation}
\psi(r,t)=\nu\left(\frac{\Sigma(r,t)}{\Sigma(r_{\odot},t)}\right)^{2(k-1)}
\left(\frac{\Sigma(r,t_{Gal})}{\Sigma(r,t)}\right)^{k-1}G^{k}_{gas}(r,t),
\end{equation}
 where $\nu$ is the efficiency of the star formation process and is
 set to be 2 Gyr$^{-1}$ for the Galactic halo  and 1
 Gyr$^{-1}$ for the disk.  $\Sigma(r,t)$ is the total
 surface mass density, $\Sigma(r_{\odot},t)$ the total surface mass
 density at the solar position, $G_{gas}(r,t)$ the surface density
 normalized to the present time total surface mass density in the disk
 $\Sigma_{D}(r,t_{Gal})$, where $t_{Gal}=14$ Gyr is the age assumed
 for the Milky Way and $r_{\odot}=8$ kpc the solar galactocentric
 distance (Reid 1993). The gas surface exponent, $k$, is set equal to
 1.5.  With these values of the parameters, the observational
 constraints, in particular in the solar vicinity, are well fitted.
 Below a critical threshold of the gas surface density
 ($7M_{\odot}pc^{-2}$) we assume no star formation.  This naturally
 produces a hiatus in the SFR between the halo-thick disk and
 the thin disk phases.

For the IMF, we use that of  Scalo (1986), constant in time and space.
$\tau_{m}$ is the evolutionary lifetime of stars as a function of their mass {\it m} (Maeder \& Maynet 1989).

The Type Ia SN rate has been computed following Greggio \& Renzini (1983)
and  Matteucci \& Greggio (1986) and it is expressed as:
\begin{equation}
R_{SNeIa}=A_{Ia}\int\limits^{M_{BM}}_{M_{Bm}}\phi(M_{B})\left[ \int\limits^{0.5}_{\mu_{m}}f(\mu)\psi(t-\tau_{M_{2}})d\mu \right]
dM_{B}
\end{equation}
where $M_{2}$ is the mass of the secondary, $M_{B}$ is the total mass of the binary
system, $\mu=M_{2}/M_{B}$, $\mu_{m}=max\left[M_{2}(t)/M_{B},(M_{B}-0.5M_{BM})/M_{B}\right]$, 
$M_{Bm}= 3 M_{\odot}$, $M_{BM}= 16 M_{\odot}$. The IMF is represented by $\phi(M_{B})$
and refers to the total mass of the binary system for the computation of the Type Ia SN rate,
$f(\mu)$ is the distribution function for the mass fraction of the secondary:
\begin{equation}
f(\mu)=2^{1+\gamma}(1+\gamma)\mu^{\gamma}  
\end{equation}
with $\gamma=2$; $A_{Ia}$ is the fraction of systems in the appropriate mass
range, which can give rise to Type Ia SN events. This quantity is fixed to
0.05 by reproducing the observed Type Ia SN rate at the present time
(Mannucci et al. 2005). Note that in the case of the Type Ia SNe
the``production matrix'' is indicated with $Q^{SNIa}_{mi}$ because of
its different nucleosynthesis contribution (for details refer to
Matteucci \& Greggio 1986 and Matteucci 2001).

The term $A(r,t)$ represents the accretion term and is defined as:
\begin{equation}
A(r,t)= a(r) e^{-t/ \tau_{H}(r)}+ b(r) e^{-(t-t_{max})/ \tau_{D}(r)}
\end{equation}
$X_{A_{i}}$ are the abundances in the infalling material, which is
assumed to be primordial, while $t_{max}=1$ Gyr is the time for the
maximum infall on the thin disk, $\tau_{H}= 0.8$ Gyr is the time scale
for the formation of the halo thick-disk and $\tau_{D} (r)$ is the
timescale for the formation of the thin disk and is a function of the
galactocentric distance (formation inside-out, Matteucci and Fran\c
cois, 1989; Chiappini et al. 2001).
 
In particular, we assume that:
\begin{equation}
\tau_{D}=1.033 r (\mbox{kpc}) - 1.267 \,\, \mbox{Gyr}
\end{equation}
Finally, the coefficients $a(r)$ and $b(r)$ are obtained  by imposing to fit  
the observed current total surface mass density in the thin disk
as a function of galactocentric 
distance given by:
\begin{equation}
\sigma(r)=\sigma_{D}e^{-r/r_{D}},
\end{equation}
where $\sigma_{D}$=531 $M_{\odot}$ pc$^{-2}$  is the central total surface mass density and
$r_{D}= 3.5$ kpc is the scale lenght.  

\section{Nucleosynthesis prescriptions}

For the nucleosynthesis prescriptions of the Fe and the other
elements (namely O, S, Si, Ca, Mg, Sc, Ti, V, Cr, Zn, Cu, Ni, Co and
Mn ), we adopted those suggested in F04.  They compared theoretical
predictions for the [el/Fe] vs. [Fe/H] trends in the solar
neighborhood for the above mentioned elements and they selected the
best sets of yields required to best fit the data.  In particular for
the yields of SNe II they found that the Woosley \& Weaver (1995)
(hereafter WW95) ones provide the best fit to the data. 
In fact, no modifications
are required for the yields of Ca, Fe, Zn and Ni as computed for solar
chemical composition. For oxygen the best results are given by the
WW95 yields computed as functions of the metallicity.  For the other
elements, variations in the predicted yields are required to best fit
the data (see F04 for details).  

 In particular, the Mg yields require some adjustments since Mg yields from massive stars are generally too low and predict a a too low Mg solar abundance. However, it is worth noting that this fact does not affect the [Mg/Fe] vs. [Fe/H] diagram if the predicted ratios are normalized to the predicted solar abundances. What matters in these diagrams are the proportions of a given elements produced in different mass ranges: Mg is mainly produced in massive stars and therefore, irrespective of the predicted Mg abundance one always obtains an overabundance of Mg relative to Fe in metal poor stars followed by a decrease of the Mg/Fe ratio at the onset of Type Ia SNe which produce the bulk of Fe. In any case, F04 increased artificially the yields of Mg from WW95 to obtain also a good solar Mg abundance and to give an indication to nucleosynthesis modelers about what might be necessary to fit the absolute Mg abundance. For the Fe-peak elements the situation is more complex and still very uncertain and in any case we do not show results for these elements in this paper. Concerning O, the best agreement with the [O/Fe] vs. [Fe/H] and the solar O abundance, as measured by Asplund et al. (2005), is obtained by adopting the original WW95 yields from massive stars as functions of metallicity. The same is not true for Fe: it was already known since the paper of Timmes \& al. (1995) that the Fe yields as functions of metallicity overestimate the solar Fe and many people use those yields divided by a factor of 2. Alternatively, one can use the yields for solar chemical composition for the whole galactic lifetime and the result is the same. This is due to the still existing uncertainties in the Fe yields.

Concerning the yields from
Type SNeIa, revisions in the theoretical yields by Iwamoto et
al.(1999) are required for Mg, Ti, Sc, K, Co, Ni and Zn to best fit
the data.  The prescription for single low-intermediate mass stars are
by van den Hoek \& Groenewegen (1997), for the case of the mass loss
parameter which varies with metallicity (see Chiappini et al. 2003a,
model5).

\section{Observational data}

We used the collection of data elaborated in the work of F04, who adopted a data sample for stars in the solar neighborhood spanning a
metallicity range from $-$4 dex to solar.  In particular, for the very
metal poor stars ([Fe/H] between $-$4 and -3 dex), the  results from the
UVES Large Program `` First Stars'' (Cayrel et al. 2003) were adopted.

For the abundances in the remaining range of [Fe/H], they adopted
already published data in the literature from various sources:
 for all the elements studied in this paper (O, Mg, Si) F04 used the data of 
 Carney et al. (1997), Nissen \& Schuster (1997), Ryan et al. (1991), Edvardsson et
al. (1993),  Matteucci
et al. (1993 and references therein).
 Moreover, concerning the Mg and Si, but not O, the collection  of F04 includes the data of Stephens (1999), Carretta et al. (2002),  McWilliam et al. (1995), Fulbright (2000), Gratton \& Sneden (1988).
 For the O, F04 also used the data of Nissen et al. (2002). Here, for the O,  we also   took into account  the data by
Bensby et al. (2004).

 All of these data are normalized
to the solar abundances of Grevesse and Sauval (1998) with the
exception of oxygen for which it is adopted the value of Asplund
et al. (2005).

For the galactic abundance gradient for the 
 oxygen we used the set of data given by Andreievsky et al. (2002a-b)
 and Luck et al. (2003) by analyzing Galactic Cepheids (see Cescutti
 at al. 2007).

\section{The galactic fountain delay model}

 In Paper I we used the Kompaneets (1960) approximation for
 the evolution of the superbubble driven by sequential supernova
 explosions. We assumed that supershells are fragmented by
 Rayleigh-Taylor instabilities, and considered each fragment as a
 cloud with an initial velocity given by the supershell velocity at
 the moment of fragmentation. Then, we followed the orbit of the
 clouds either ballistically (the cloud is subject only to the
 gravitational potential of the Galaxy) or taking into account viscous
 interaction between the extra-planar gas halo and the cloud.

In Tables 2-4 reported in  Paper I, are summarized the
results for fragmentation times ($t_{final}$) and the velocities
($v_{n}$) of the supershells in the direction perpendicular to the
galactic plane at those times for initial throwing coordinates fixed
at 4 kpc, 8 kpc and 12 kpc. In the Table \ref{tempimedi} here we report
the average time $<t_{total}>$ necessary to create a cloud from a
supershell plus the time necessary to the cloud to fall back into the
Galactic disk:
\begin{equation}
<t_{total}>=t_{final}+<t_{orbit}>,
\end{equation}
 
where $<t_{orbit}>$ is the average time required for a cloud to
return to the galactic disk once it leaves the supershell.  These
average values are calculated in the case of OB associations with 10,
50, 100, 500 SNe at 3 galactocentric distances: 4 kpc, 8 kpc, 12 kpc.
In Table \ref{delaymax} we show the maximum delay time  $t_{total}$ , at a
fixed initial galactocentric distance of ejection.

\begin{table}[h!]
\caption{The average time  $<t_{total}>$ [Myr]. }
\label{tempimedi}
\begin{center}
\begin{tabular}{cccc}
  \hline\hline

\noalign{\smallskip}
& 4 kpc   & 8 kpc &12 kpc\\
\noalign{\smallskip}

  \hline
\noalign{\smallskip}
  10 SNe& 43  & 53 &75\\
% &  &  \\
 50 SNe&36&54&87\\
% &  &  \\
 100 SNe& 36 & 57& 96\\
% &  &  \\
 500 SNe& 38 & 75& 133\\
% &  &  \\ 

 \hline
 \end{tabular}
\end{center}
\end{table}

\begin{table}[h!]
\caption{ The maximum $t_{total}$ as a function of the galactocentric distance. }
\label{delaymax}
\begin{center}
\begin{tabular}{cccc}
  \hline\hline

\noalign{\smallskip}
& 4 kpc   & 8 kpc &12 kpc\\
\noalign{\smallskip}

  \hline
\noalign{\smallskip}
  Maximum delay& 48 Myr  &114 Myr& 245 Myr\\

% &  &  \\ 

 \hline
 \end{tabular}
\end{center}
\end{table}

\begin{figure}
\includegraphics[width=0.45\textwidth]{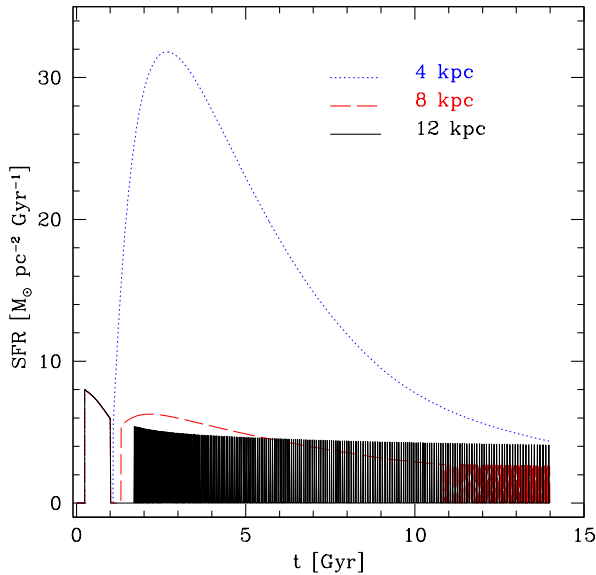}
\caption{The star formation rate expressed in $M_{\odot}$ pc$^{-2}$
  Gyr$^{-1}$ as a function of the galactic time (in Gyr), as predicted
  by the two infall model of Fran\c{c}ois et al. (2004) at 4 (dotted
  line), 8 (dashed line) and 12 kpc (solid line). }
\label{SFR}
\end{figure}

\begin{table}[h!]
\caption{Beginning of the star formation in the disk after the halo-thick disk transition ($t_{SFR}$).}
\label{SFRinizio}
\begin{center}
\begin{tabular}{cccc}
  \hline\hline

\noalign{\smallskip}
& 4 kpc   & 8 kpc &12 kpc\\
\noalign{\smallskip}

  \hline
\noalign{\smallskip}
  Beginning of SF in the disk& 1.1 Gyr  & 1.3 Gyr& 1.7 Gyr\\

% &  &  \\ 

 \hline
 \end{tabular}
\end{center}
\end{table}

As mentioned in Sect.1, a Galactic fountain can affect the
chemical evolution of the Galaxy only by means of a time-delay due to
the non-negligible time taken by fountains to orbit around and fall
back into the Galaxy. We implemented this delay in the chemical
evolution model of F04  only during the thin disk phase.   In fact, 
the dynamical effect of a galactic
  fountain requires the presence of a stratified ISM, which arises
  only after the end of the halo phase. 
   In the past, several authors (i.e.  Mac Low et al. 1989 and Tenorio-Tagle 1996), studied the supershell evolution produced by a sequential explosion of Type II SNe. They emphasized  that a break-out event, necessary for a galactic fountain, requires that  the OB associations must sit on a {\it plane stratified disk} where the density decreases along the $z$ direction. In the work of Mac Low et al. (1989) a scale height of 100 pc for the ISM was considered. This is the reason why OB associations existed at early times during the formation of the halo but did not produce galactic fountains. Galactic fountains could have formed perhaps during the thick disk phase. However, by computing a fountain with a scale height of 1 kpc (typical of the thick disk), the fountains start at $\simeq3$ kpc, i.e. well within the halo. At this height the stratification is much weaker and the ISM temperature is high, therefore the sound speed is high and the shock can be easily dissipated through sound waves. Moreover the thick disk phase lasts for a very short time (see Fig. \ref{SFR}). We can then conclude that (hypothetical) thick disk fountains would have a negligible impact on the chemical evolution of the Galaxy.

 Therefore, we consider the galactic fountain delay only
  for ejecta from massive stars (with mass $>$ 8 $M_{\odot}$) born after
  the halo-thick disk transition. 
  In Fig.  \ref{SFR} we show the predicted star formation
rate (SFR) expressed in $M_{\odot}$ pc$^{-2}$ Gyr$^{-1}$ as a function
of the galactic time (Gyr) using the model of F04. We note that the
SFR is the same for all galactocentric distances during the halo
phase. The effect of the two infall model in presence of
the threshold in the star formation is clearly shown: the star formation halts
during the halo-thin disk transition and the duration of the gap
varies as a function of galactocentric distance. At the solar
neighborhood the gap is $\simeq$ 0.3 Gyr. This gap seems to be
observed in the abundance ratio distribution as shown in Gratton et
al. (2000) and Fuhrmann (1999). In Table \ref{SFRinizio} we report the
times at different galactocentric distances at which the SFR in the
thin disk begins ($t_{SFR}$) in the F04 model.  
 Ejecta of intermediate-mass stars and Type Ia SNe (which are most probably not clumped in OB associations) are not subject to the galactic fountain delay.

As shown in Table \ref{delaymax} a single superbubble can produce a
delay in the mixing up to $\sim$ 250 Myr, but the average values are
smaller, spanning  the range between 40 - 140 Myr (Table
\ref{tempimedi}). In the simulation we considered different values up
to 1 Gyr. It is possible that the Milky Way formed super-star clusters
during more intense phases of star formation, as observed in M82 (Melo
et al. 2005). A delay of 1 Gyr can then be obtained if the number of
massive stars in a cluster is $\sim 10^4$, which is consistent with
the mass of the largest super-star clusters observed in M82. In any
case, we considered this long delay as an extreme case.

\section{The metal cooling delay model}

Roy \& Kunth (1995) studied the dispersal and mixing of oxygen in the
ISM of gas-rich galaxies. Due to a variety of hydrodynamical processes
operating at scales ranging from 1 pc to greater than 10 kpc in the
ISM, there are difficulties in estimating the dispersal and
mixing timescale: $(i)$ on large galactic scales (1 $\geq\ l\ \geq$ 10 kpc),
turbulent diffusion of interstellar clouds in the shear flow of
galactic differential rotation is able to wipe out azimuthal O/H
fluctuations in less than $10^9$ yrs; $(ii)$ at the intermediate scale
(100 $\geq\ l\ \geq$ 1000 pc), cloud collisions and expanding
supershells driven by evolving associations of massive stars,
differential rotation and triggered star formation will re-distribute
and mix gas efficiently in about $10^8$ yrs; $(iii)$ at small scales
(1 $\geq\ l\ \geq$ 100 pc), turbulent diffusion may be the dominant
mechanism in cold clouds, while Rayleigh-Taylor and Kelvin-Helmhotz
instabilities quickly develop in regions of gas ionized by massive
stars, leading to full mixing in $\leq 2 \times 10^6$ yrs.  Malinie et
al. (1993) claimed that due to chemical inhomogeneities observed for
example in the age-metallicity relation in the solar neighborhood,
re-mixing and star formation may be delayed by $10^{8-9}$ yr.

Following the work of Thomas et al. (1998) we relaxed the IMA in the 
chemical evolution model by splitting the gas component into two
different phases (cold and warm).  The gas in the interstellar medium
is heated by SN events and the stars cannot form in this ``warm
gas phase''; only after the gas cools stars can form and then pollute the
ISM.  We implemented the two gas phases simply by means of a delay in the
chemical enrichment. In this case: i) all stellar masses (both Type II
 and Type Ia SNe) and ii) both halo and disk stars, are affected by
this delay.

\section{Results}

In this section we discuss our results. First of all we  show
how a galactic fountain delay can affect the chemical evolution model,
then we present the results for the metal cooling delay model.

We show the model predictions for O, Mg, Si and Fe obtained in particular
for the relations [el/Fe] versus [Fe/H] compared with the
observational data and try to put constraints on the maximum possible
delay.

\subsection{Results for the  galactic fountain delay model}

We have taken into account different values for the delay originated by
galactic fountain events from sequential SN explosions: 0.1 Gyr, 0.2
Gyr, 0.5 Gyr, 1 Gyr. We tested the effect at different galactocentric
distances: 4, 8, 12 kpc.

\begin{figure}
\includegraphics[width=0.45\textwidth]{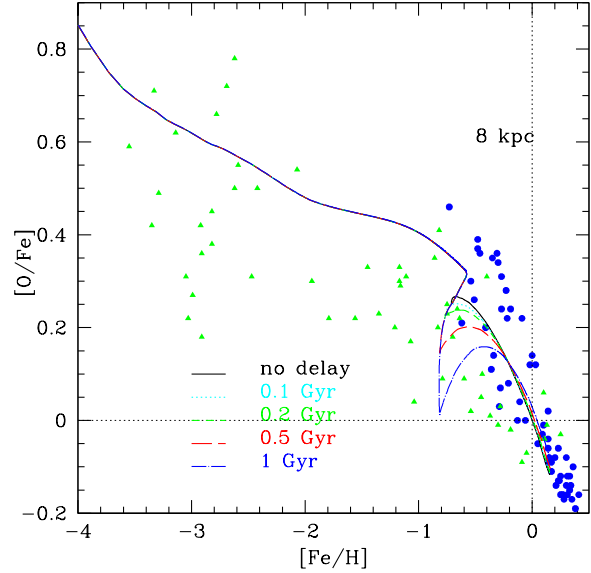}
\caption{The abundance ratio [O/Fe] as a function of [Fe/H] in the
  solar neighborhood. We compared the Fran\c{c}ois et al. (2004) model
  (solid line) with our galactic fountain delay models: dotted line
  (0.1 Gyr), short dashed  line (0.2 Gyr), long  dashed
  line (0.5 Gyr), dashed dotted line (1 Gyr). The data are taken
  from the collection used in Fran\c{c}ois et al. (2004) (filled green triangles)  and from Bensby et al. (2005) (filled blue circles).  }
\label{Ofemetdep}
\end{figure}

Concerning the solar neighborhood, in Table \ref{solar} we report the
solar abundances by mass as predicted by the F04 model and the delay
models compared to the solar values of Asplund et al. (2005).  In
Fig. \ref{Ofemetdep} we show the results for [O/Fe] versus [Fe/H] in
the solar neighborhood compared to the observational data described in
Sect 4. Our results are normalized to the ``best model'' of F04 values
at 9.5 Gyr (the age of the formation of the solar system). As we
  can see from this figure and from Table \ref{solar}, the solar
abundances predicted by our models are substantially
in agreement with those derived by Asplund et al. (2005).
 The standard model with no delay shows a discontinuity at $\sim -0.8$ dex, and this is due to the gap in the SFR discussed before. In fact, during the gap O stops being produced whereas Fe keeps being produced by Type Ia SNe. The effect of the delay due to the fountains is therefore to increase this effect by increasing the duration of the gap.

\begin{table*}[ht!]
\caption{Mass fractions of several elements yielded by the
  Fran\c{c}ois et al. (2004) model and the delay model at the galactic
  age of 9.5 Gyr compared to the solar values of Asplund et al. (2005).}
\label{solar}
\begin{center}
\begin{tabular}{|c|c|c|c|c|c|c|}
  \hline\hline
% & &\multicolumn{5}{|c|}{}\\
  & $X_{\odot oss} $ &\multicolumn{5}{|c|}{ $X_{\odot th} $}\\
%\noalign{\smallskip}
%& &\multicolumn{5}{|c|}{}\\
  \hline
%\noalign{\smallskip}
    & Asplund et al. (2005) &Fran\c{co}is et al. (2004)  &0.1 Gyr &0.2 Gyr &0.5 Gyr &1 Gyr \\ 

  \hline
%\noalign{\smallskip}
%    &  &  & & & &\\ 
  He  &   2.55  $\times 10^{-1}$ & 2.52 $\times 10^{-1}$&  2.52 $\times 10^{-1}$  &  2.52 $\times 10^{-1}$ &2.52 $\times 10^{-1}$  &2.52 $\times 10^{-1}$  \\
%   &  &  & & & &\\ 
  C  & 2.21 $\times 10^{-3}$   & 2.03 $\times 10^{-3}$&2.05 $\times 10^{-3}$ &2.06 $\times 10^{-3}$ &2.11 $\times 10^{-3}$ & 2.17 $\times 10^{-3}$ \\
%    &  &  & & & &\\ 
     N  &    6.32 $\times 10^{-4}$  & 9.73 $\times 10^{-4}$ &9.75 $\times 10^{-4}$ &9.77 $\times 10^{-4}$ &9.81 $\times 10^{-4}$ &9.85 $\times 10^{-4}$\\
%    &  & & & & & \\ 
  O  & 5.48 $\times 10^{-3}$    & 5.27 $\times 10^{-3}$ & 5.31 $\times 10^{-3}$ & 5.35 $\times 10^{-3}$ & 5.46 $\times 10^{-3}$ & 5.62 $\times 10^{-3}$ \\
%    &  &  & & & &\\ 

  Mg  & 6.17  $\times 10^{-4}$   &6.66 $\times 10^{-4}$ &6.70 $\times 10^{-4}$ &6.74 $\times 10^{-4}$ & 6.86 $\times 10^{-4}$& 7.03 $\times 10^{-4}$\\
%    &  &  & & & &\\ 

  Si  &  6.80$\times 10^{-4}$   &8.14 $\times 10^{-4}$ &8.17 $\times 10^{-4}$ &8.20 $\times 10^{-4}$ &8.30 $\times 10^{-4}$ &8.43 $\times 10^{-4}$ \\
%    &  &  & & & &\\ 
 S  &  3.31 $\times 10^{-4}$   & 3.95$\times 10^{-4}$ & 3.97$\times 10^{-4}$ & 3.98$\times 10^{-4}$ &4.03$\times 10^{-4}$ &4.09$\times 10^{-4}$\\
%    &  & & & & &\\ 
  Ca  &  6.13 $\times 10^{-5}$    & 5.40 $\times 10^{-5}$ &5.42 $\times 10^{-5}$ &5.44 $\times 10^{-5}$ &5.50 $\times 10^{-5}$ &5.58 $\times 10^{-5}$ \\
%    &  &  & & & &\\ 
 Fe  &  1.18  $\times 10^{-3}$  &  1.14 $\times 10^{-3}$ & 1.15 $\times 10^{-3}$ &1.15 $\times 10^{-3}$ &1.15 $\times 10^{-3}$ &1.16 $\times 10^{-3}$\\
%    &  &  & & & &\\ 

 \hline
 \end{tabular}
\end{center}
\end{table*}

  In the description of the delay model we emphasized that galactic fountains
  can be seen only in disk stars. The typical delay, due to the fact
  that massive stars are clumped in OB associations, is 0.1
  Gyr. Referring to  Fig. \ref{Ofemetdep} we present a first strong
  result: {\it a delay of 0.1 Gyr produces a negligible effect on the
    chemical evolution of the Galaxy in the solar neighborhood.}
 We note that the data show a
   spread in the halo-thick disk transition phase ([Fe/H]$\simeq$-0.77).  This spread can be
   explained in terms of a combination of: i) a threshold in the star
   formation, ii) a delay in the chemical enrichment from the massive
   stars due to galactic fountain effect. Moreover, 
we note 
 that the maximum possible delay,
  in order not to break the agreement with data,
   must be lower than 1 Gyr.

\begin{figure}
\includegraphics[width=0.45\textwidth]{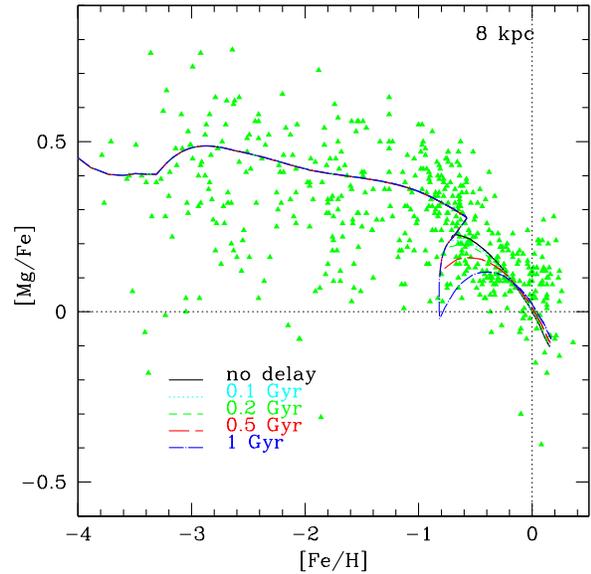}
\caption{The abundance ratio [Mg/Fe] as a function of [Fe/H] in the
  solar neighborhood. Notation as in Fig. \ref{Ofemetdep}. }
\label{mg8}
\end{figure}

\begin{figure}
\includegraphics[width=0.45\textwidth]{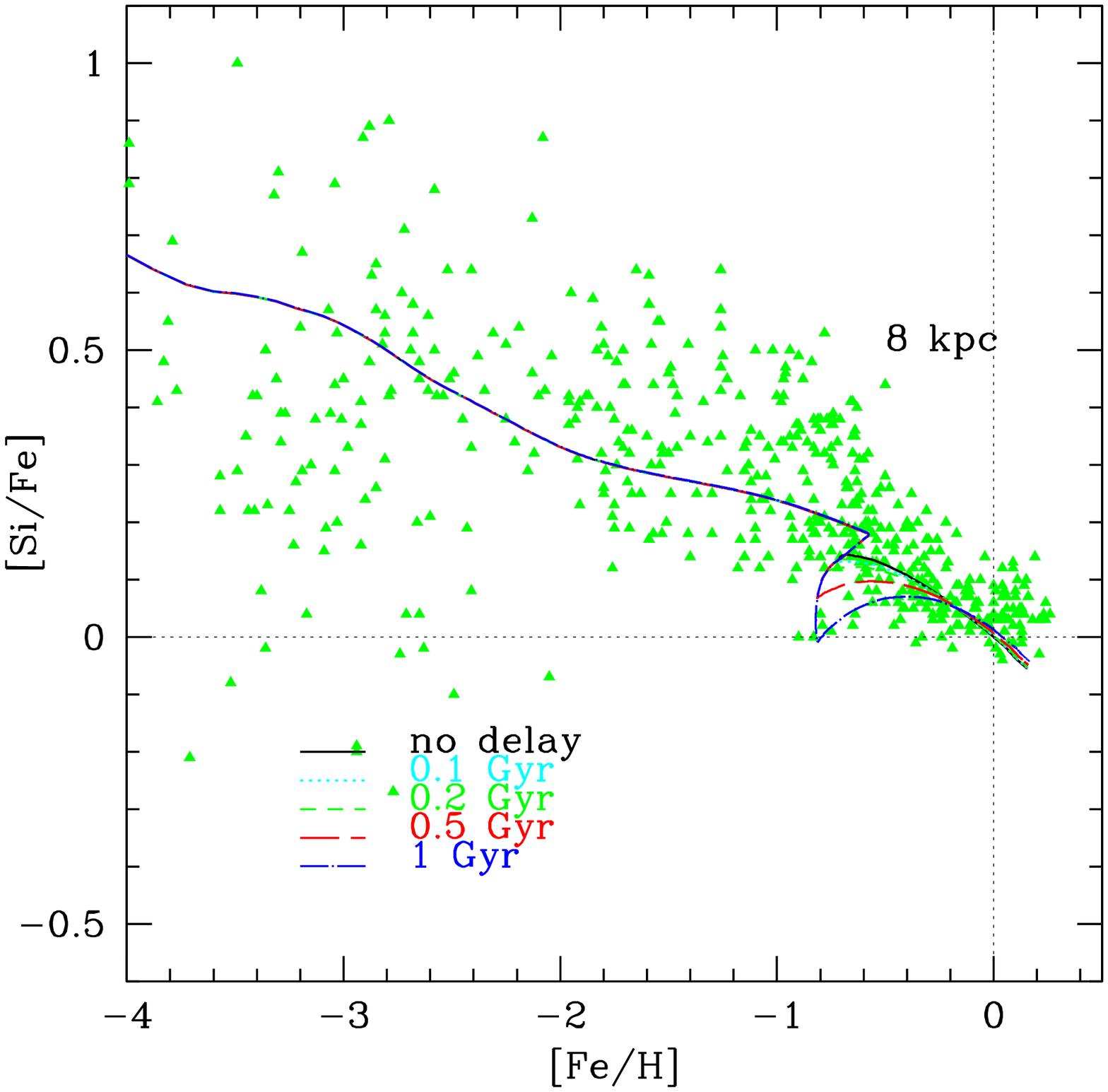}
\caption{The abundance ratio [Si/Fe] as a function of [Fe/H] in the solar neighborhood.  Notation as in Fig. \ref{Ofemetdep}.}
\label{si8}
\end{figure}

 The same effect is shown in Fig. \ref{mg8} and in Fig. \ref{si8} where we analyzed the behavior of such a delay in a chemical evolution
model for other two $\alpha$ elements: Mg and Si.
     
 \begin{figure}
\includegraphics[width=0.45\textwidth]{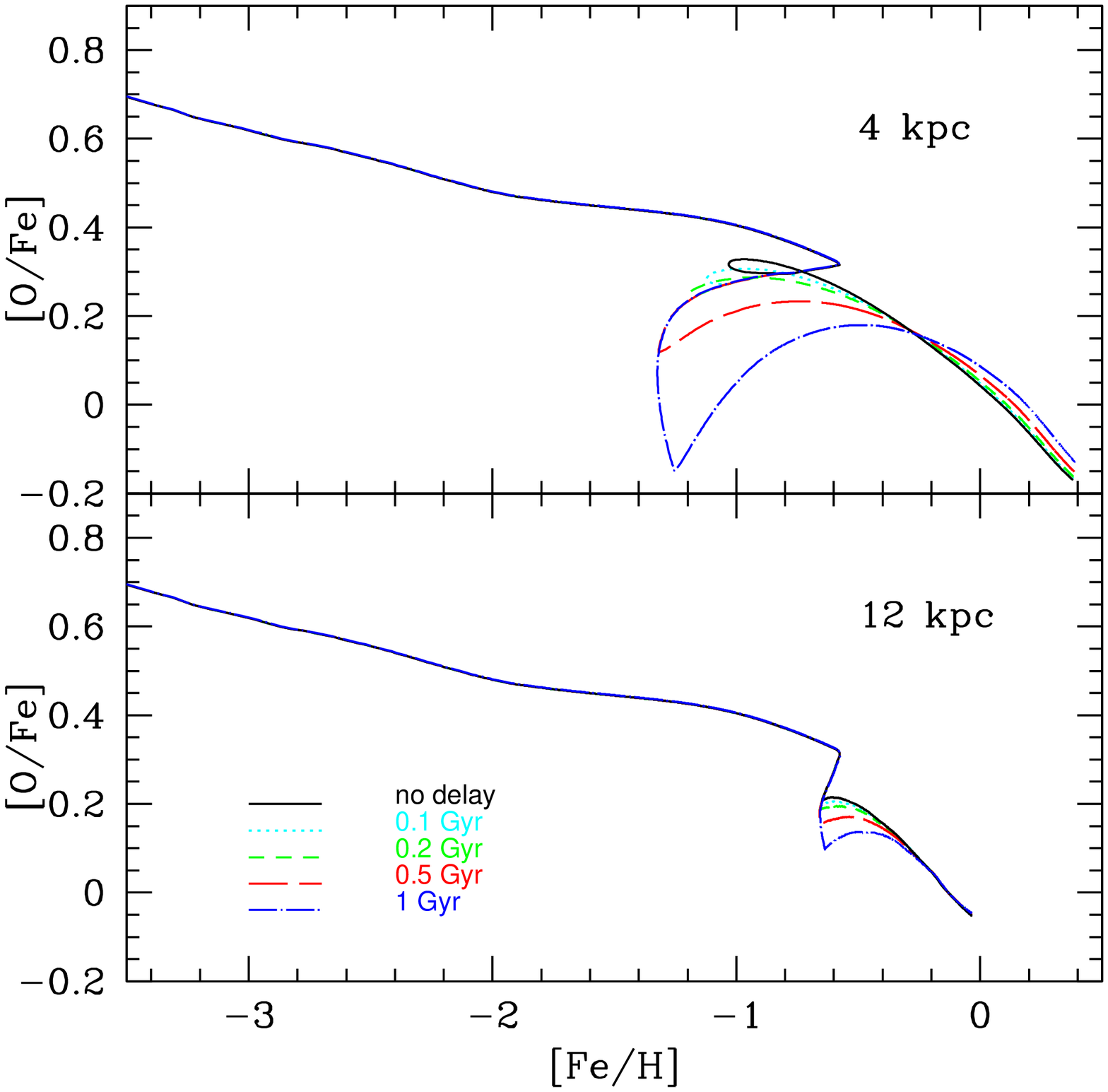}
\caption{The abundance ratio [O/Fe] as a function of [Fe/H] at 4 and
  12 kpc.  Notation as in Fig. \ref{Ofemetdep}.}
\label{Oinsieme}
\end{figure}

\begin{figure}
\includegraphics[width=0.45\textwidth]{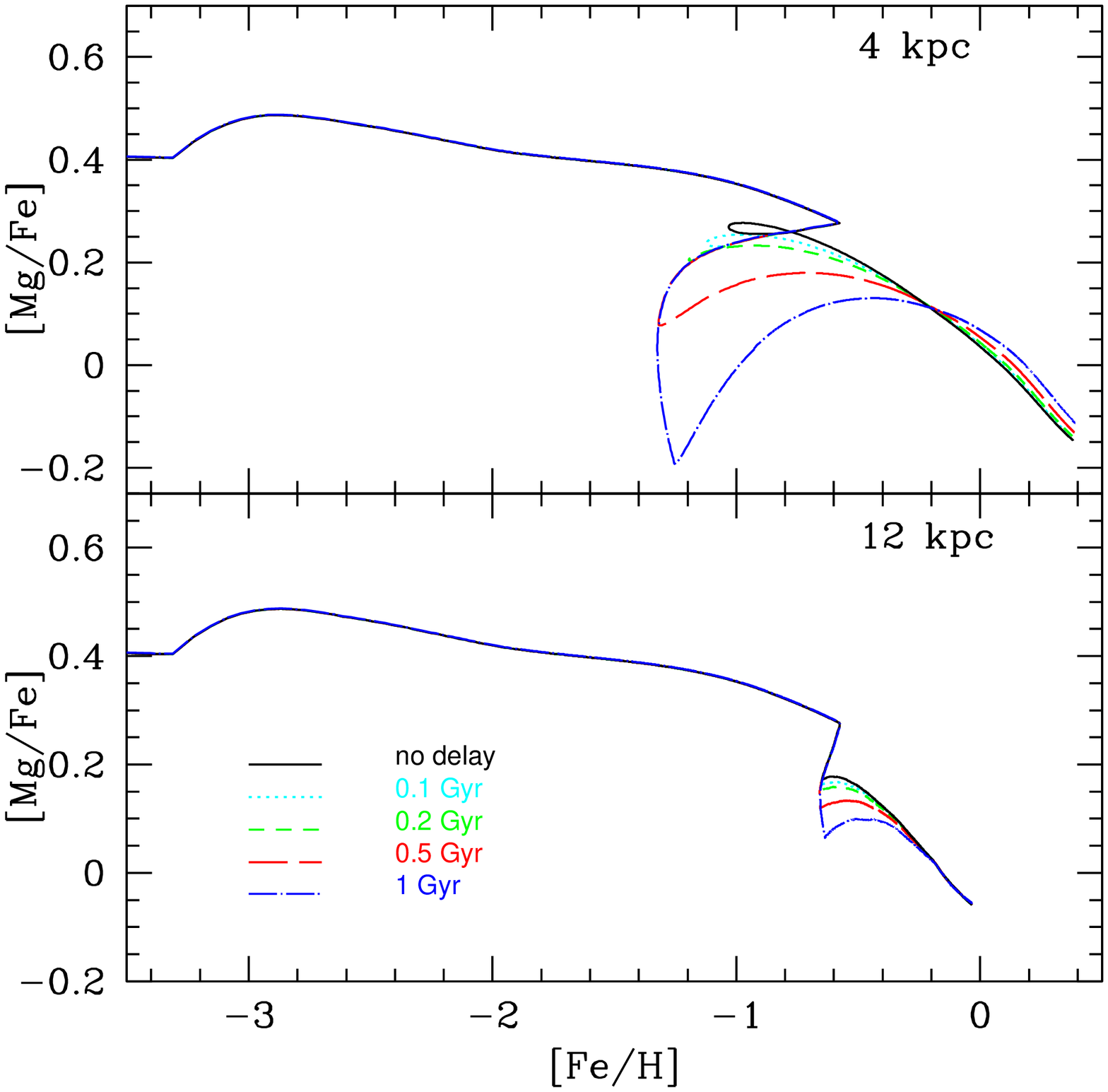}
\caption{The abundance ratio [Mg/Fe] as a function of [Fe/H] at 4 and 12 kpc. Notation as in Fig. \ref{Ofemetdep}.}
\label{Mginsieme}
\end{figure}

\begin{figure}
\includegraphics[width=0.45\textwidth]{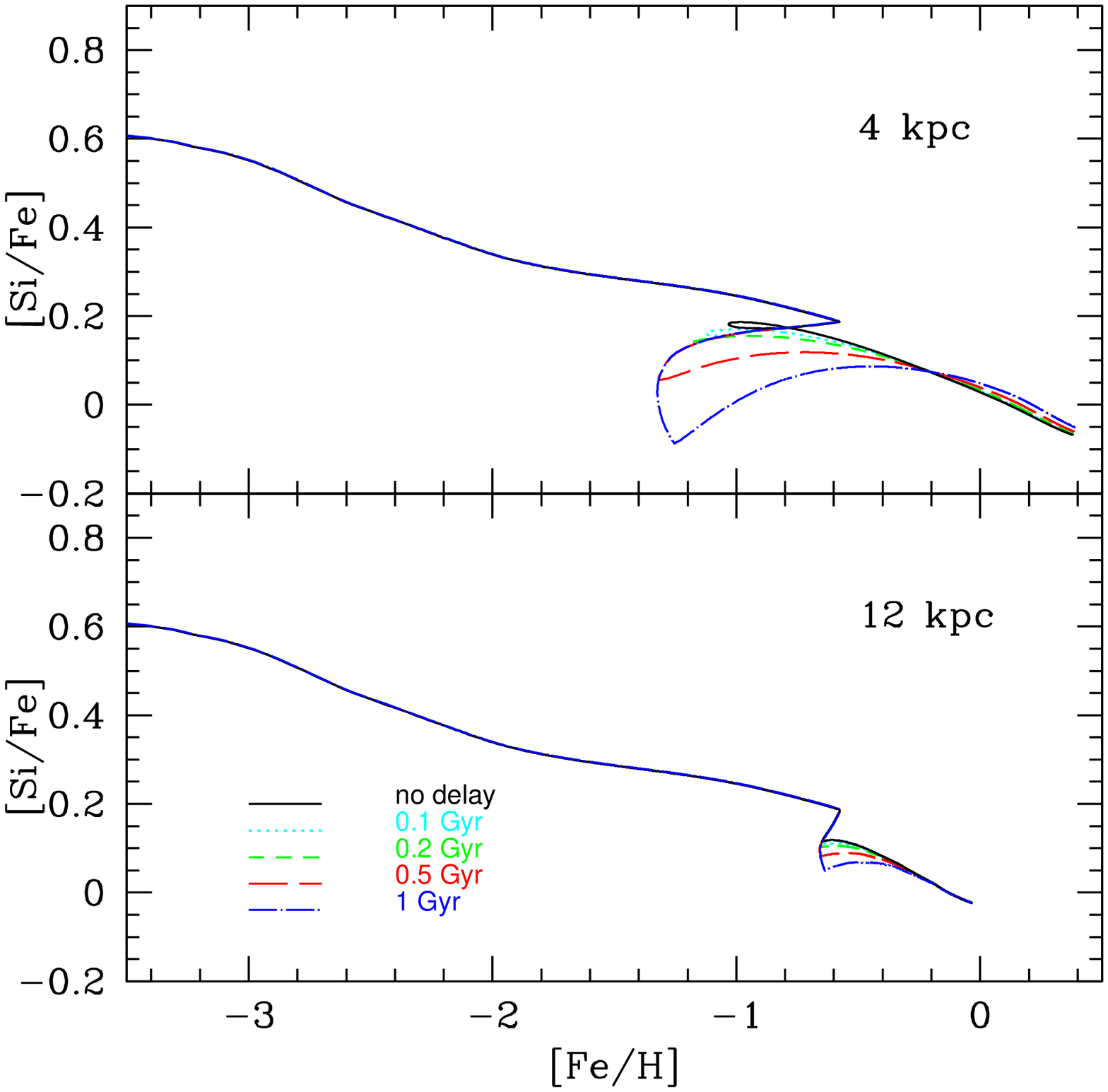}
\caption{The abundance ratio [Si/Fe] as a function of [Fe/H] at 4 and 12 kpc Notation as in Fig. \ref{Ofemetdep}.}
\label{Siinsieme}
\end{figure}

\begin{figure}
\includegraphics[width=0.45\textwidth]{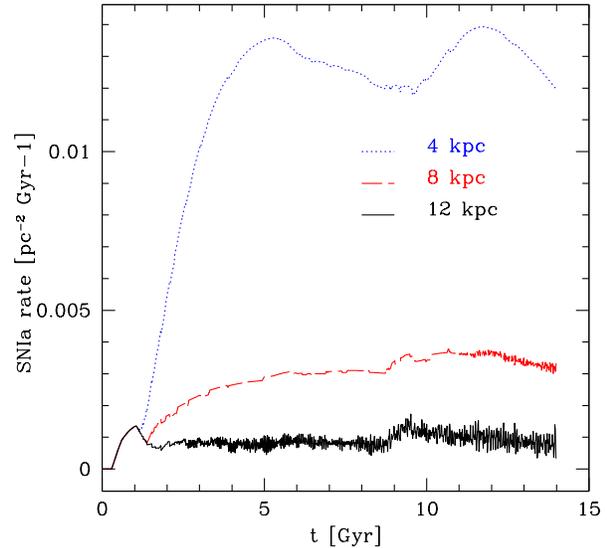}
\caption{The Type Ia SN rate expressed in pc$^{-2}$ Gyr$^{-1}$ as a
  function of the galactic time (Gyr), as predicted by the two infall
  model of Fran\c{c}ois et al. (2004) at 4, 8 and 12 kpc. }
\label{SNIA}
\end{figure}

\begin{figure}
\includegraphics[width=0.45\textwidth]{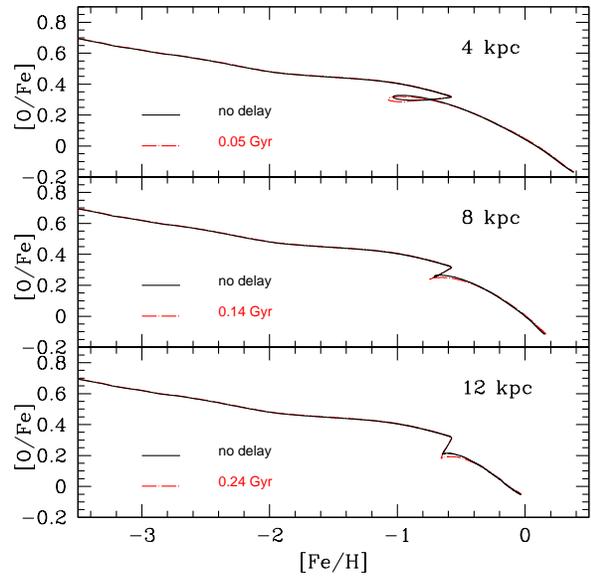}
\caption{The abundance ratio [O/Fe] as a function of [Fe/H] at 4, 8  and
  12 kpc for relative maximum  delays as reported in Tab \ref{delaymax}.  }
\label{max}
\end{figure}

\begin{figure}[ht!]
\includegraphics[width=0.45\textwidth]{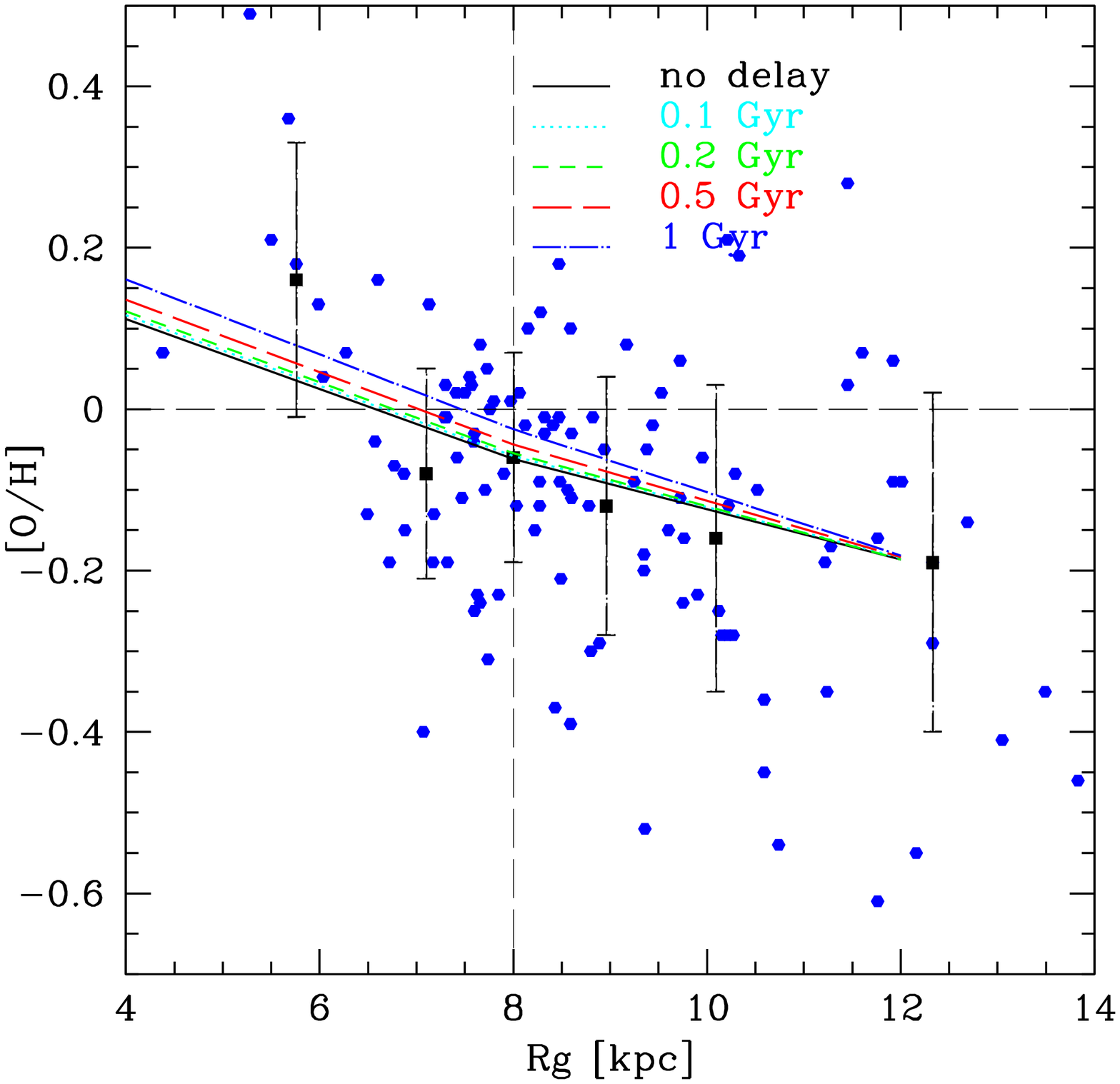}
\caption{Predicted and observed O abundance gradients in the
  galactocentric distance range 4-14 kpc. The data points are from
  Cepheids. The large squares with error bars represent average of the
  points with their errors. The black solid
  line is the Fran\c{c}ois et al. (2004) model normalized to the mean
  value of the bin centered at 8 kpc (see Cescutti et al. 2007). }
\label{gradient}
\end{figure}

It is also interesting to test the fountain delay at different
galactocentric distances. In Figs. \ref{Oinsieme}, \ref{Mginsieme} and
\ref{Siinsieme} we report our results for O, Mg and Si,
respectively. In each figure we plot the [el/Fe] vs. [Fe/H] at 4 and
12 kpc galactocentric distances. We note that the results are strongly
dependent on the galactocentric distance. In fact, as explained
before, the drop of the [el/Fe] quantity in the [el/Fe] vs. [Fe/H] is
due to the iron ejected by the Type Ia SNe. 
 The discontinuity  predicted at 4 kpc is larger than that predicted at larger galactocentric distances. This effect is due to the higher specific Type Ia SN rate at 4 kpc than at larger distances. The specific Type Ia SN rate is the SNIa rate per unit mass of gas and it increases toward the Galactic center, the reason being that while the Type Ia SN rate is a factor of 8 higher than that in the solar vicinity, because of the higher SFR, the mass of gas is higher by only a factor of 2 relative to the solar region. Therefore, the effect of the pollution by Type Ia SNe during the gap is enhanced.

Again in  Figs. \ref{Oinsieme}, \ref{Mginsieme} and \ref{Siinsieme}
we note that, at a fixed galactocentric distance, the effect of the
galactic fountain delay depends also on the considered element: the Si
which is also produced by Type Ia SNe in a non negligible amount
shows a smaller drop of the [Si/Fe] quantity compared to O and Mg.

 Even if the galactic fountain effect in the [el/Fe] values is
  larger at 4 kpc because of the higher rate of Type Ia SNe, the
  delays expected at 4 kpc are smaller with respect to outward
  galactocentric distances (as shown Tabs \ref{tempimedi} and
  \ref{delaymax}), as a consequence of the higher gravitational pull. In
  Fig. \ref{max} we show the [O/Fe] vs [Fe/H] at 4, 8 and 12 kpc
  considering the maximum delays in each radius. In the case of a delay
  produced by a single OB association, the effect of galactic
  fountains is negligible.

 Finally we explored the effects of galactic fountains on the
 abundance gradient for the oxygen. In Fig. \ref{gradient} we plot the
 [O/H] as a function of the galactocentric distance. We note that
 the average delay due to galactic fountains  is longer than 0.1 Gyr
 only for a OB association composed of 500 SNe at 12 kpc (Table
 \ref{tempimedi}); this event is very unlikely, then the delay of
 0.1 Gyr can be considered like a average upper limit for a delay
 produced by a single OB association at a galactocentric distance
  in the range 4-12 kpc. Referring to Fig.\ref{gradient} we
 can conclude that {\it the time delay produced by a Galactic fountain
   originated by an OB association has a negligible effect on the
   abundance gradient in the Galaxy disk}.

\subsection{Results for the metal cooling delay model}
\begin{figure}
\includegraphics[width=0.45\textwidth]{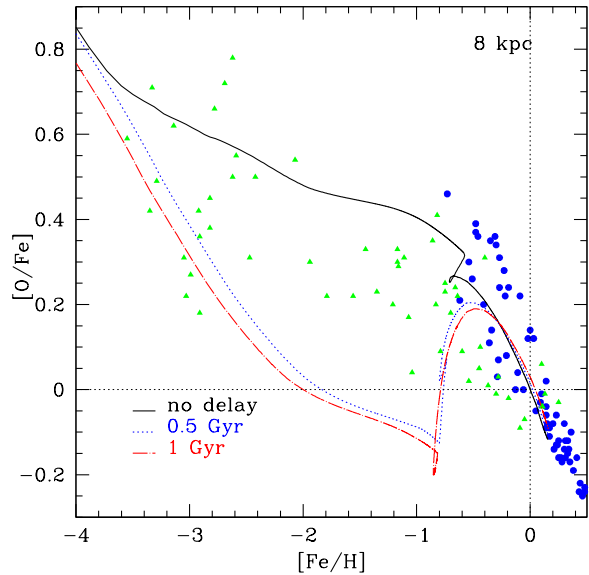}
\caption{The abundance ratio [O/Fe] as a function of [Fe/H] in the
  solar neighborhood predicted by our metal cooling delay model. For
  the O we considered the metallicity dependent yields of WW95. Solid
  line: model without delay; dotted line and dashed dotted line: 0.5
  and 1 Gyr delays, respectively. Symbols as in Fig. \ref{Ofemetdep}.}
\label{Ocaldo}
\end{figure}

\begin{figure}
\includegraphics[width=0.45\textwidth]{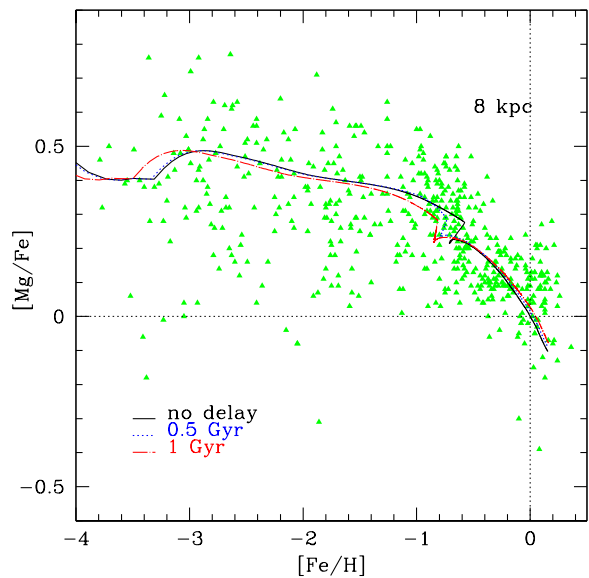}
\caption{The abundance ratio [Mg/Fe] as a function of [Fe/H] in the
  solar neighborhood predicted by our metal cooling delay model. Notation as in Fig.\ref{Ocaldo}.}
\label{Mgcaldo}
\end{figure}

\begin{figure}
\includegraphics[width=0.45\textwidth]{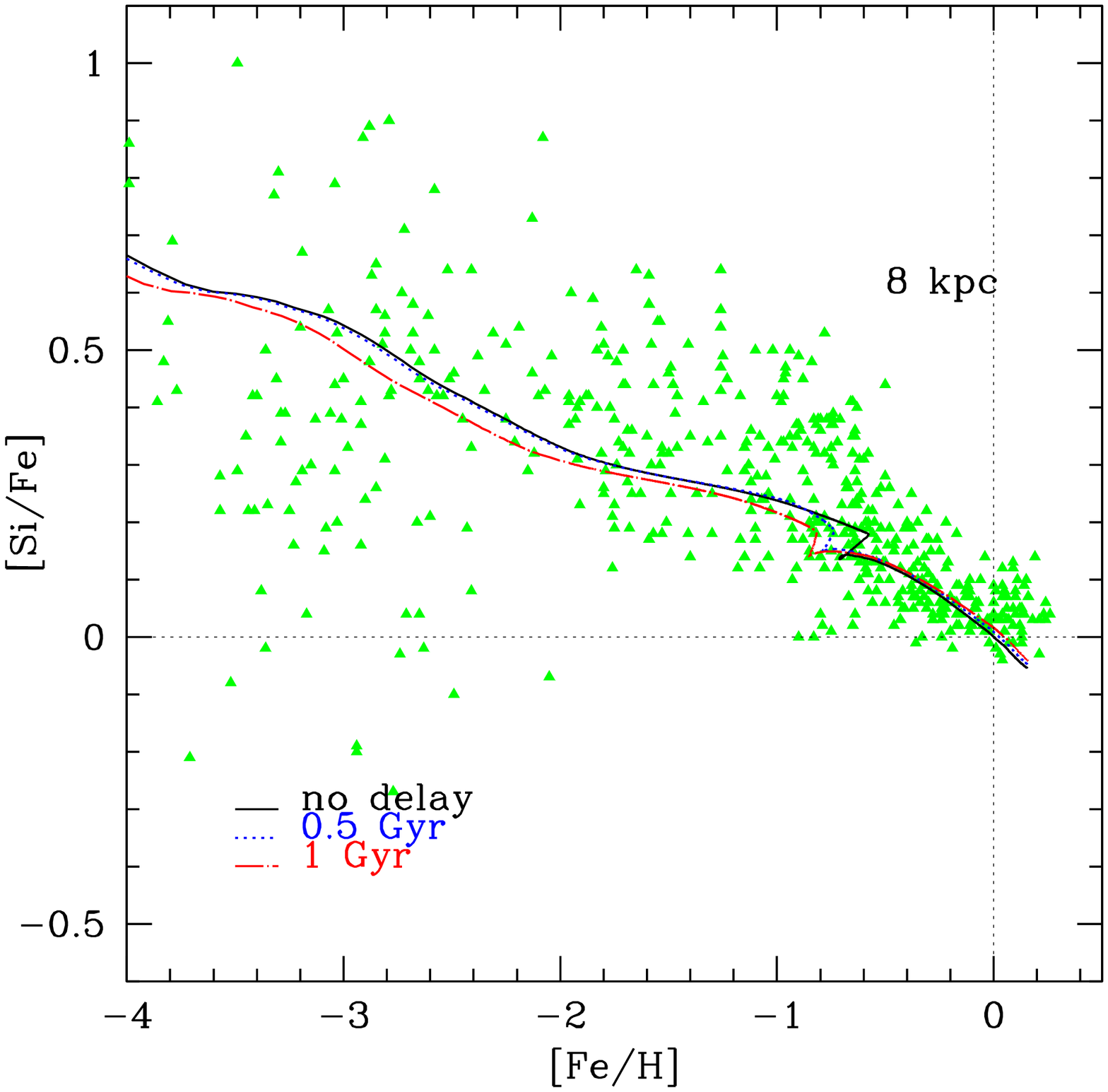}
\caption{The abundance ratio [Si/Fe] as a function of [Fe/H] in the
  solar neighborhood predicted by our  the metal cooling delay model. Notation as in Fig.\ref{Ocaldo}.}
\label{Sicaldo}
\end{figure}

In this section we present the results for the metal cooling delay
model. Referring to Malinie et al. (1993) we considered mixing delays 
of 0.5 Gyr and 1 Gyr. In Figs. \ref{Ocaldo},
\ref{Mgcaldo} and \ref{Sicaldo} we show the effect of the metal
cooling predicted by our model using the [el/Fe] vs. [Fe/H] plots. The work of Thomas et al. (1998)  already discussed the effect
of a two- phase gas model relative to Mg. Referring to Fig.
\ref{Mgcaldo} we obtain a similar result, but the effect of our delay
in the mixing is smaller. Also for the Si we have a similar behavior
as can be seen in the Fig.  \ref{Sicaldo}. The differences between ours and Thomas et al.'s results are probably due to the different adopted yields for massive stars and IMF. In fact, Thomas et al. adopted Thielemann et al. (1996) yields and the Salpeter (1955) IMF.

\begin{figure}
\includegraphics[width=0.45\textwidth]{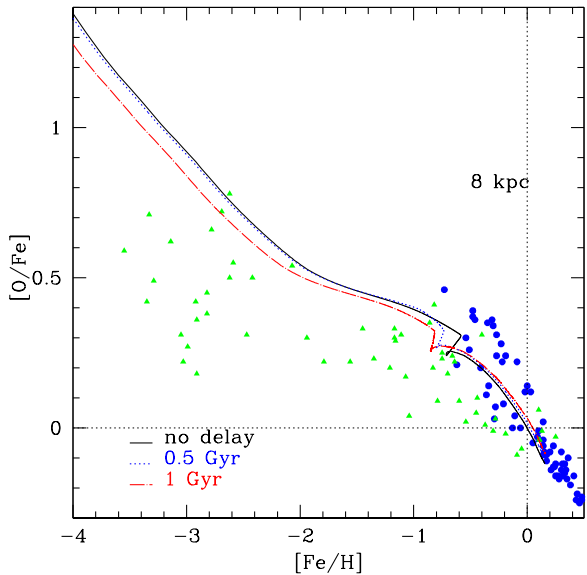}

\caption{The abundance ratio [O/Fe] as a function of [Fe/H] in the
  solar neighborhood predicted by our metal cooling delay model. For the O
  we considered the solar yields of WW95. Notation as in Fig.\ref{Ocaldo}. }
\label{Ocaldonomet}
\end{figure}

\begin{figure}
\includegraphics[width=0.45\textwidth]{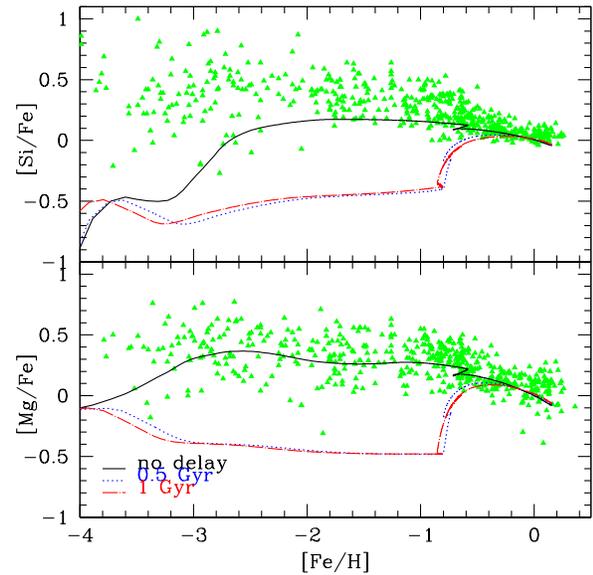}
\caption{The abundance ratio [el/Fe] as a function of [Fe/H] for the
  Si and Mg in the solar neighborhood predicted by our metal cooling
  delay model. For the Si and Mg we considered the metal dependent
  yields of WW95. Notation as in Fig.\ref{Ocaldo}. }
\label{MgSimetdep}
\end{figure}

 A different result is obtained for the [O/Fe] vs [Fe/H] reported
 in Fig.  \ref{Ocaldo}. In this case, we find that the delay in
 the mixing leads to big differences in terms of chemical evolution
 models. We see that in the halo-thick disk phase, the predicted
 [O/Fe] values are much smaller than for the model with the IMA. The
 reason for this resides in the fact that in the model of F04 the
 yields used for the O are the WW95 metallicity dependent ones and in
 this case the delay induces a situation where the yields for Z=0,
 lower than those computed for the other metallicities, act for a longer time, 
 thus
 producing the large depression in the [O/Fe] ratio. To demonstrate
 that this result is solely due to the choice of nucleosynthetic
 yields, we plot in Fig. \ref{Ocaldonomet} the metal cooling delay
 model results using, for the oxygen, the yields at solar metallicity
 given by WW95. As we can see from this figure, in this case the
 models with delay do not differ considerably compared to the
 reference model and the oxygen behaves similarly to the other
 $\alpha$-elements. In Fig. \ref{MgSimetdep} we report the [el/Fe]
 vs. [Fe/H] relations for the Mg and Si using for these two elements
 the metallicity dependent yields of WW95. As expected, the [el/Fe]
 ratios show a large depression in the halo-thick disk phase as in
 Fig. \ref{Ocaldo} due to the choice of metallicity dependent
 yields.  Therefore we can conclude that the metallicity-dependent
 yields of WW95 are not compatible with a delayed enrichment of the
 halo.

\begin{figure}
\includegraphics[width=0.45\textwidth]{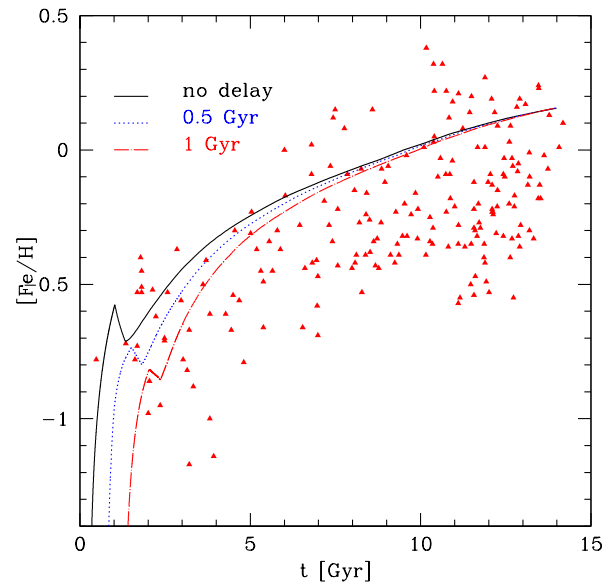}
\caption{The abundance ratio [Fe/H] as a function of the galactic time [Gyr] in the solar neighborhood predicted by our metal cooling
  delay model. The data are taken from  Ramirez et al. (2007) (filled red triangles). Notation as in Fig.\ref{Ocaldo}.}
\label{Fet}
\end{figure}

\begin{figure}
\includegraphics[width=0.45\textwidth]{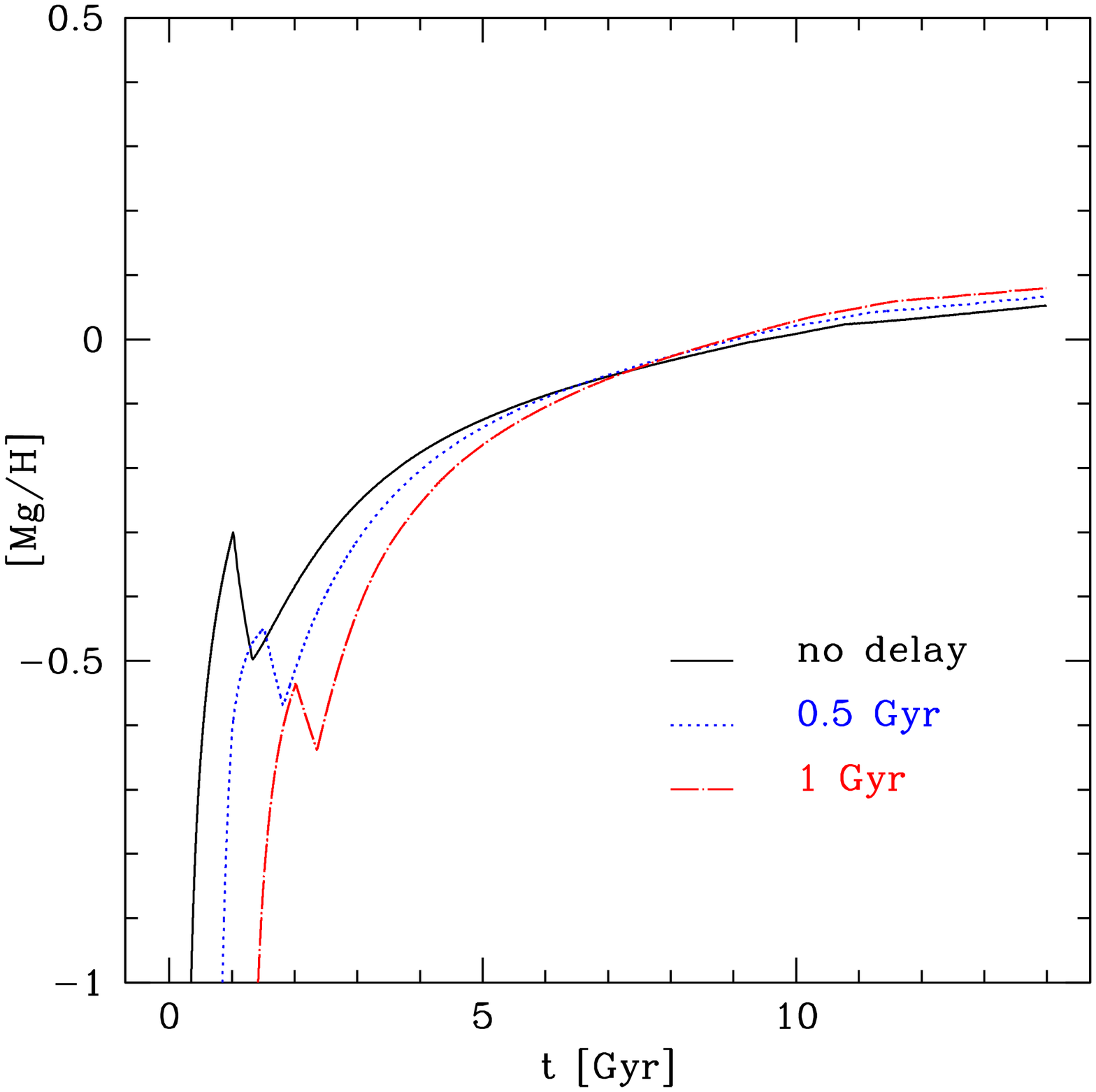}
\caption{The abundance ratio [Mg/H] as a function of the galactic time [Gyr] in the solar neighborhood predicted by our metal cooling
  delay model. Notation as in Fig.\ref{Ocaldo}.}
\label{Mgt}
\end{figure} 

 Finally, in Figs. \ref{Fet} and \ref{Mgt} we report age-metallicity relations in the solar neighborhood for [Fe/H] and [Mg/H] respectively, predicted by our metal cooling delay model using the stellar yields adopted in F04. Concerning the evolution of the [Fe/H] we compare our results with the observational data of Ramirez et al. (2007). As one can see, the effects of the delays are more visible in the age-metallicity relations than in the abundance ratios versus [Fe/H], and the reason is that in the abundance ratios these differences are partly erased. We conclude that all delays considered are compatible with the observations owing to  the very large  data  spread for the age metallicity relation, but also  that the age-metallicity relation is not a good constraint for chemical models.

\section{Conclusions}

In this paper we have studied the relaxation of IMA in chemical
evolution models by means of the effects of a delay in the chemical
enrichment produced by galactic fountain events and by means of a
delay in the mixing due to chemical inhomogeneities in the disk
(our metal cooling delay model). The main purpose of this work  was
testing the results of Paper I in a chemical evolution
model. In the present paper we tested the ``delay'' effect due to the finite
and not negligible time for a cloud spend to fall back onto the
disk.

Our main conclusions can be summarized as follows:   

\begin{itemize}

\item In the solar neighborhood we showed that the average delay
  produced by the galactic fountains has a negligible effect on 
  the chemical evolution for all the $\alpha$ elements we studied.

\item In the [el/Fe] versus [Fe/H] relations, the main feature of the galactic
  fountain is an enhancement of the drop in the [el/Fe] ratios occurring because of the two infall scenario. In fact, the drop in the standard model of F04 is due to the halt of the SFR which produces  a halt in the production of the $\alpha$-elements while the Fe from Type Ia SNe continues to be produced. The galactic fountain delay has the effect of increasing the period during which there is no pollution from Type II SNe. 

\item {Results produced by the model with a galactic fountain delay
  of 1 Gyr are not compatible with observational data. On the other
  hand, any delay $<$ 1 Gyr could be acceptable, given the observed
  spread in the data.}

\item {\it The time delay produced by a Galactic fountain originated
  by a OB association has a negligible effect on the abundance
  gradients in the Galaxy disk}.

\item The metal cooling delay model with  the assumed delays 
  (which are those suggested in the literature), has a very small effect
  on the chemical evolution in the solar neighborhood if yields not
  depending on metallicity are used.

\item On the other hand, in the case of the metal dependent yields of WW95,
  the results differ substantially from
  the reference model of F04 and do not fit the observations.

\end{itemize}

\begin{acknowledgements}
We thank the referee for the enlightening suggestions.
We also  thank F. Calura, G. Cupani and A. Saro for many useful discussions.We also thank I.J. Danziger for reading the manuscript. We acknowledge financial support from the PRIN 2007 MUR Prot. N. 2007JJC53X$\_$001.  

\end{acknowledgements}

\end{document}